\documentclass[twocolumn,aps,prl,amsmath,amssymb]{revtex4}
\usepackage{graphicx}
\usepackage{dcolumn}
\usepackage{bm}
\usepackage[dvips]{color}
\usepackage{epstopdf}
\DeclareGraphicsExtensions{.pdf,.eps,.png,.jpg,.mps}
\begin{document}

\author{D. De Fazio$^1$}
\author{I. Goykhman$^1$}
\author{M. Bruna$^1$}
\author{A. Eiden$^1$}
\author{S. Milana$^1$}
\author{D. Yoon$^1$}
\author{U. Sassi$^1$}
\author{M. Barbone$^1$}
\author{D. Dumcenco$^2$}
\author{K. Marinov$^2$}
\author{A. Kis$^2$}
\author{A. C. Ferrari$^1$}
\email{acf26@eng.cam.ac.uk}
\affiliation{$^1$Cambridge Graphene Centre, University of Cambridge, Cambridge CB3 0FA, UK}
\affiliation{$^2$Electrical Engineering Institute, Ecole Polytechnique Federale de Lausanne, Switzerland}

\title{High Responsivity, Large-Area Graphene/MoS$_2$ Flexible Photodetectors}

\begin{abstract}
We present flexible photodetectors (PDs) for visible wavelengths fabricated by stacking centimetre-scale chemical vapor deposited (CVD) single layer graphene (SLG) and single layer CVD MoS$_2$, both wet transferred onto a flexible polyethylene terephthalate substrate. The operation mechanism relies on injection of photoexcited electrons from MoS$_2$ to the SLG channel. The external responsivity is 45.5A/W and the internal 570A/W at 642nm. This is at least two orders of magnitude higher than bulk-semiconductor flexible membranes and other flexible PDs based on graphene and layered materials. The photoconductive gain is up to $4\times10^5$. The photocurrent is in the 0.1-100$\mu$A range. The devices are semi-transparent, with just 8$\%$ absorption at 642nm and work stably upon bending to a curvature of 6cm. These capabilities and the low voltage operation ($<$ 1V) make them attractive for wearable applications.
\end{abstract}
\maketitle
\section{Introduction}
Modern electronic and opto-electronic systems such as smart phones, smart glasses, smart watches, wearable devices and electronic tattoos increasingly require ultra-thin, transparent, low-cost and energy efficient devices on flexible substrates\cite{AkinNC2014}. The rising demand for flexible electronics and optoelectronics requires materials which can provide a variety of electrical and optical functionalities, with constant performance upon application of strain\cite{Ryha2010}. A wide range of optoelectronic devices on flexible substrates have been reported to date, such as photodetectors (PDs)\cite{YuanAPL2009,LiuOE2013}, light emitting diodes (LEDs)\cite{ParkS2009}, optical filters\cite{QianAPL2008}, optical interconnects\cite{ChenNM2011,BosmPTL2010}, photovoltaic devices\cite{ShahNP2010,YoonNC2011} and biomedical sensors\cite{KoN2008,KimS2011}.

Major challenges in the development of flexible optoelectronic devices stem from the limitations associated with the high stiffness of bulk semiconductors\cite{MacmJMS1972, BlakJAP1982}. In the case of flexible PDs, the current approaches primarily rely on thin ($\mu$m-thick) semiconductor membranes\cite{YangAPL2010,YuanAPL2009} and compound semiconductor nanowires (NWs)\cite{LeeMEMS2012,LiuOE2013,ChenJMCC2014,YuJMCC2014}, mainly because of their ability to absorb light throughout the whole visible range (0.4-0.7$\mu$m) and the possibility to adapt their fabrication techniques from rigid to plastic, or deformable substrates\cite{AkinNC2014}.

One of the key parameters for PDs characterization is the responsivity. This is defined as the ratio between the collected photocurrent ($I_{ph}$) and the optical power. The responsivity is named external ($R_{ext}=I_{ph}/P_o$)\cite{Sze2006} or internal ($R_{int}=I_{ph}/P_{abs}$)\cite{Sze2006}, whenever the incident ($P_o$) or absorbed ($P_{abs}$) optical power is used at the denominator. Since not all incident photons are absorbed, i.e. $P_{abs}<P_{in}$, then $R_{int}$ is typically larger than $R_{ext}$\cite{Sze2006}.

In flexible PDs, $R_{ext}$ up to$\sim0.3A/W$ was reported for crystalline semiconductor membranes (InP, Ge)\cite{YangAPL2010,YuanAPL2009} with integrated p-i-n junctions, showing photocurrent up to$\sim100\mu A$, with$\sim30\%$ degradation upon bending at a radius $r_b\sim$3cm\cite{YangAPL2010}. PDs made of a single semiconductor NW on flexible substrates\cite{LeeMEMS2012, LiuOE2013, ChenJMCC2014, YuJMCC2014} demonstrated $R_{ext}$ up to$\sim10^5 A/W$, for $r_b$ down to 0.3cm\cite{LiuOE2013}. Yet, these provide limited $I_{ph}$ in the order of $nA$\cite{LiuOE2013, ChenJMCC2014, YuJMCC2014} up to less than $1\mu A$\cite{LeeMEMS2012}. For flexible devices exploiting NW-arrays by drop-casting\cite{LiuOE2013, ChenJMCC2014, YuJMCC2014}, rather than based on single-NWs, $R_{ext}$ degrades significantly from$\sim10^5 A/W$ to the $mA/W$ range\cite{LiuOE2013, ChenJMCC2014, YuJMCC2014}, due to photocurrent loss at multiple junctions in the NW network\cite{LiuOE2013, ChenJMCC2014, YuJMCC2014}.

Graphene and related materials (GRMs) have great potential in photonics and optoelectronics\cite{BonaNP2010, FerrN2015, SunACS2010, KoppNN2014}. A variety of GRM-based devices have been reported, such as flexible displays\cite{KimN2009}, photovoltaic modules\cite{BaugNN2014,PospNN2014}, photodetectors\cite{KoppNN2014, XiaNN2009, KisNN2013}, optical modulators\cite{LiuN2011}, plasmonic devices\cite{ChenN2012, FeiN2012, JuNN2011, YanNN2012, EchtNC2011}, and ultra-fast lasers\cite{SunACS2010}. Heterostructures, obtained by stacking layers of different materials were also explored\cite{FerrN2015, KoppNN2014}, e.g. in photovoltaic\cite{FurcNL2014} and light emitting devices\cite{WithNM2015}. Refs.\citenum{RoyNN2013, ZhanSR2014} reported SLG/MoS$_2$-based PDs made of mechanically exfoliated\cite{RoyNN2013} or CVD grown\cite{ZhanSR2014} materials, transferred on Si/SiO$_2$ rigid substrates. These reached back-gate dependent $R_{int}\sim$10$^8$A/W for optical intensities$<$0.1pW/$\mu$m$^{2}$.

GRM based flexible PDs have been reported for visible light ($0.4-0.7\mu m$\cite{WithNL2014,FinnJMCC2014}) using materials produced by liquid phase exfoliation (LPE)\cite{HernNN2008,BonaMT2012} of graphene and transition metal dichalcogenides (TMDs)\cite{WithNL2014,FinnJMCC2014}. In Ref.\citenum{WithNL2014}, a flexible PD on polyethylene terephthalate (PET) was fabricated by sandwiching a LPE dispersion of WS$_2$ between a LPE graphene film serving as a back electrode and a chemical vapor deposited (CVD) graphene top electrode, with $R_{ext}\sim$0.1mA/W. This value is orders of magnitude lower compared to other flexible PDs\cite{YangAPL2010,YuanAPL2009,LeeMEMS2012, LiuOE2013, ChenJMCC2014, YuJMCC2014}. The lower $R_{ext}$ in LPE-based devices is attributed to non-efficient inter-flake charge transfer\cite{KingACS2010,TorrNN2014}, resulting in limited conductivity\cite{KingACS2010,TorrNN2014} and poor collection of photo-generated carriers at the outer metal electrodes\cite{WithNL2014}. Similarly, the inefficient charge transfer affected $R_{ext}$ (in the $nA/W$ range) in Ref.\citenum{FinnJMCC2014}, where a PD was fabricated with LPE MoS$_2$ as absorber and LPE graphene as top electrode.

Here we take advantage of the mechanical properties of layered materials to demonstrate flexible gate-controlled SLG/MoS$_2$ PDs for visible wavelengths with $R_{ext}$ of tens of A/W and optical transparency$>80\%$. The devices are assembled by stacking on a PET substrate a centimetre-scale CVD SLG on a CVD-grown single layer MoS$_2$ (1L-MoS$_2$). In this configuration, 1L-MoS$_2$ acts as visible light absorber, while SLG is the conductive channel for $I_{ph}$ flow\cite{ZhanSR2014,RoyNN2013}. We show that $R_{ext}$ increases either by promoting carrier injection from 1L-MoS$_2$ to SLG using polymer electrolyte gating, or by increasing the source-drain voltage. We get $R_{ext}$ up to$\sim45A/W$ applying a 1V bias with $I_{ph}\sim$tens $\mu A$. These values are at least two orders of magnitude higher than those reported in semiconductor membranes devices\cite{YangAPL2010,YuanAPL2009}, semiconductor NW arrays\cite{LeeMEMS2012, LiuOE2013, ChenJMCC2014,YuJMCC2014} and GRM-based\cite{WithNL2014,FinnJMCC2014} flexible PDs operating in the visible range\cite{YangAPL2010,YuanAPL2009,LeeMEMS2012, LiuOE2013,ChenJMCC2014,YuJMCC2014,WithNL2014,FinnJMCC2014}. This $R_{ext}$ is achieved in devices with$\sim82\%$ transparency, twice that reported for semiconductor membrane devices\cite{YangAPL2010}. We get $R_{int}\sim570$A/W for$\sim$0.1nW/$\mu$m$^{2}$ at 642nm, similar to SLG/MoS$_2$ PDs\cite{ZhanSR2014,RoyNN2013} on rigid substrate operating at the same optical power level. This shows that SLG/MoS$_2$ heterostructures on PET retain their photodetection capabilities. Upon bending, our devices have stable performance for r$_b$ down to$\sim6$cm. This is comparable to r$_b$ measured in semiconductor membranes PDs\cite{YangAPL2010,YuanAPL2009}, which show lower ($<0.3A/W$) responsivities\cite{YangAPL2010,YuanAPL2009}. Although our r$_b$ is one order of magnitude larger than for flexible single NWs\cite{LeeMEMS2012, LiuOE2013, ChenJMCC2014, YuJMCC2014}, the latter had at least three orders of magnitude smaller device areas ($<5\mu m^2$)\cite{LeeMEMS2012, LiuOE2013, ChenJMCC2014, YuJMCC2014} compared to our PDs ($>0.2mm^2$). Given the responsivity, flexibility, transparency and low operation voltage (below $1V$), our PDs may be integrated in wearable, biomedical and low-power opto-electronic applications\cite{LeeMEMS2012,KoN2008,KimS2011}.
\section{Results and discussion}
\begin{figure}
\centerline{\includegraphics[width=90mm]{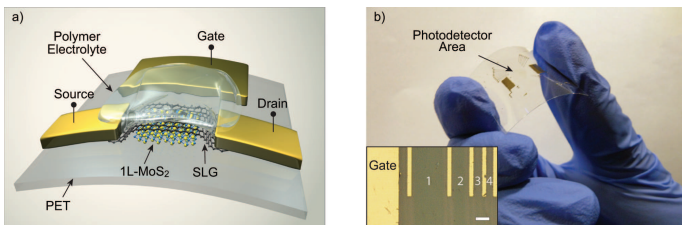}}
\caption{a) Schematic SLG/MoS$_2$ flexible PD, side-gated with a polymer electrolyte. b) Picture of a typical PD, showing transparency and flexibility. (Inset) Optical image of 4 PDs with different channel lengths and common side gate electrode. Scale bar is 200$\mu$m.}
\label{fig.1}
\end{figure}
\begin{figure*}
\centerline{\includegraphics[width=160mm]{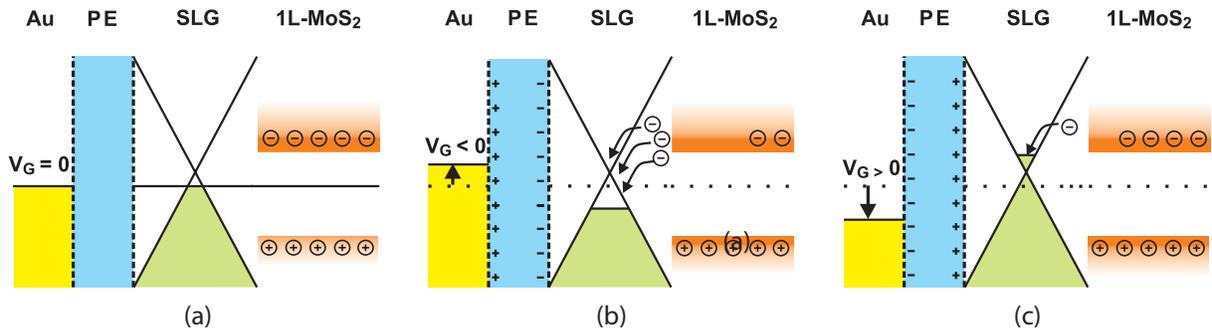}}
\caption{Schematic band diagram of polymer electrolyte (PE) gated SLG/1L-MoS$_2$ PD at a) zero, b) negative, c) positive $V_{GS}$}
\label{fig.2}
\end{figure*}
Fig.\ref{fig.1} plots a schematic drawing of our PD. The device consists of a 1L-MoS$_2$ absorber covered by a SLG channel, clamped between source and drain electrodes. We chose PET as a flexible substrate due to its$\sim$90\% transparency in the visible range\cite{FaraOAM2011} and ability to withstand solvents\cite{MartPNAS2013} (e.g acetone and isopropyl alcohol) commonly used in the transfer processes of 2d materials grown by CVD (e.g. transfer of SLG grown on Cu)\cite{BaeNN2010}. The SLG/1L-MoS$_2$ heterostructure is gated using a polymer electrolyte\cite{DasPRB2009,DasNN2008}.

The operation principle of our devices is depicted in Fig.\ref{fig.2}. For energy bands alignment, the electron affinity of 1L-MoS$_2$ and the Dirac point of SLG are assumed to be$\sim$4-4.2eV\cite{ChoiNC2013, DasNL2013} and$\sim$4.6eV\cite{YuNL2009, ShanPRL2005}, respectively. We also assume SLG to be initially p-doped (Fig.\ref{fig.2}a), as reported in previous works involving SLG transferred on PET substrates\cite{KimNL2010,LeeNL2012}. In thermodynamic equilibrium, $E_F$ is constant across the device and is located below the Dirac point. During illumination, part of the photo-generated electrons would be injected from the 1L-MoS$_2$ conduction band into the p-doped SLG\cite{ZhanSR2014,RoyNN2013}, leaving behind uncompensated photogenerated holes. The latter would act as an additional positive $V_{GS}$ to the SLG channel, seen as a shift of the charge neutrality point ($V_{CNP}$) to more negative voltages. In p-doped SLG, the injected electrons from 1L-MoS$_2$ would occupy energy states above $E_F$ (Fig.\ref{fig.2}b), thus reducing the concentration of holes and decreasing the PD current. Electron injection can be further promoted by gating. When negative $V_{GS}$ is applied, higher p-doping of the SLG channel would induce a stronger electric field at the SLG/1L-MoS$_2$ interface\cite{ZhanSR2014}, thus favoring electron transfer from 1L-MoS$_2$ (Fig.\ref{fig.2}b). Hence, for negative $V_{GS}$, $R_{ext}$ is expected to increase, due to injection of more electrons and consequent more pronounced PD current reduction. The opposite should happen for positive $V_{GS}$, where the gate-induced negative charge in SLG would reduce the p-doping and shift $E_F$ towards the Dirac point. In this case, the photogenerated electrons in 1L-MoS$_2$ would experience weaker electric fields, becoming less attracted by the SLG channel. As a result, we expect $R_{ext}$ to decrease. When positive $V_{GS}$ is high enough, $E_F$ would cross the Dirac point and SLG would become n-doped (Fig.\ref{fig.2}c). Thus, only weak electron injection from 1L-MoS$_2$ would be possible if $E_F$ in SLG remains below the 1L-MoS$_2$ conduction band; the transferred electrons increase free-carriers concentration in the n-doped channel, hence minor increments of $R_{ext}$ and $I_{ph}$ are expected.

Our devices are built as follows. 1L-MoS$_2$ is epitaxially grown by CVD on c-plane sapphire substrates\cite{DumcACS2015}. These are annealed at 1000$^{\circ}$C in air for 1 hour after consecutive cleaning by acetone/isopropyl alcohol/deionized (DI) water. They are then placed face-down above a crucible containing$\sim$5mg MoO$_3$ ($\geq$99.998\% Alfa Aesar). This is loaded into a 32mm outer diameter quartz tube placed in a split-tube three-zone furnace. A second crucible containing 350mg sulfur ($\geq$ 99.99\% purity, Sigma Aldrich) is located upstream from the growth substrates. Ultrahigh-purity Ar is used as carrier gas at atmospheric pressure. The procedure is: ramp the temperature to 300$^{\circ}$C with 200sccm Ar flow, set to 300$^{\circ}$C for 10mins, ramp to 700$^{\circ}$C (50$^{\circ}$C/min increase temperature rate) with 10sccm Ar flow, set at 700$^{\circ}$C for 10 min, cool to 570$^{\circ}$C with 10sccm of Ar, increase the gas flow to 200sccm and open the furnace for rapid cooling\cite{DumcACS2015}. SLG is grown on 35$\mu$m Cu foil, following the process described in Ref.\citenum{BaeNN2010}. The substrate is annealed in hydrogen atmosphere (H$_2$, 20sccm) up to $1000^{\circ}$C for 30 minutes. Then, 5sccm CH$_4$ is added to initiate growth\cite{LiS2009, BaeNN2010}. The sample is then cooled in vacuum (1mTorr) to room temperature and removed from the chamber.

\begin{figure*}
\centerline{\includegraphics[width=160mm]{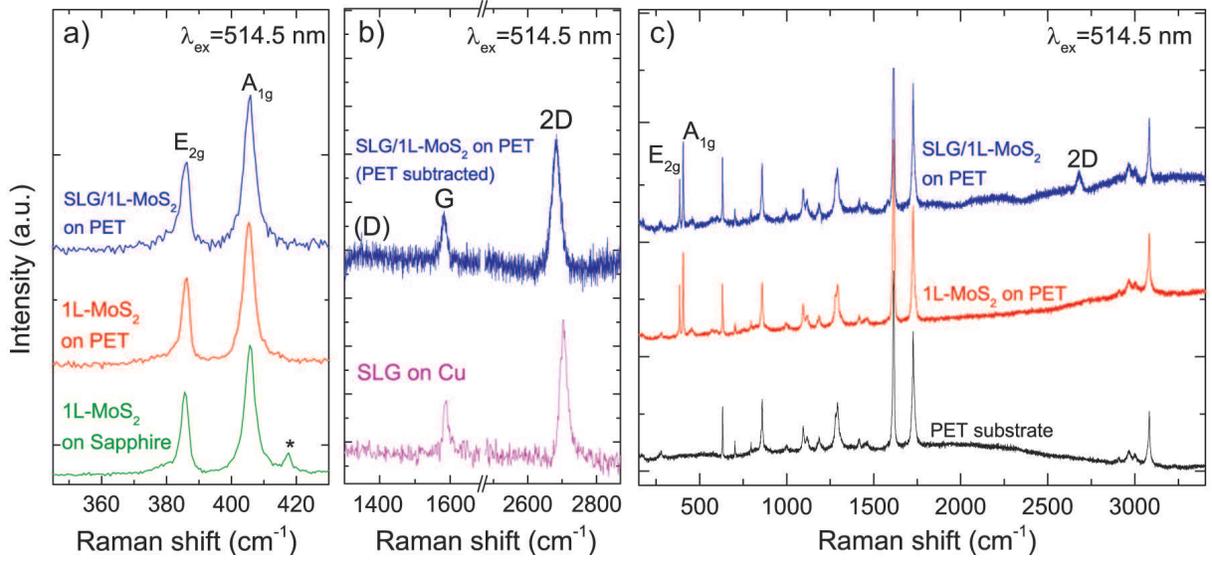}}
\caption{(a) Raman spectra at 514nm for 1L-MoS$_2$ on sapphire, 1L-MoS$_{2}$ on PET, and SLG/1L-MoS$_2$. (b) Comparison at 514nm of the Raman spectra of as-grown SLG on Cu (magenta curve) with SLG/1L-MoS$_2$ after transfer on PET. (c) Raman spectra at 514nm of PET substrate (black curve), 1L-MoS$_{2}$ on PET (red curve) and SLG/1L-MoS$_2$ on PET (blue curve).}
\label{fig.3}
\end{figure*}

Prior to assembling the SLG/MoS$_2$ stack, the quality and uniformity of MoS$_2$ on sapphire and SLG on Cu are inspected by Raman spectroscopy and photoluminescence (PL), using a Horiba Jobin Yvon HR800 spectrometer equipped with a 100X objective. The laser power is kept below 100$\mu$W (spot size $<1\mu$m in diameter) to avoid possible heating effects or damage. Fig.\ref{fig.3}a (green curve) plots the Raman spectrum of CVD MoS$_{2}$ on sapphire for 514nm excitation. The peak at$\sim385$cm$^{-1}$ corresponds to the in-plane (E$_{2g}^1$) mode\cite{VerbPRL1970,WietPRB1971}, while that at$\sim$404cm$^{-1}$, is the out of plane (A$_{1g}$) mode\cite{VerbPRL1970,WietPRB1971}, with full width at half maximum FWHM(E$_{2g}^1$)=2.5 and FWHM(A$_{1g}$)=3.6cm$^{-1}$, respectively. The E$_{2g}^1$ mode softens, whereas the A$_{1g}$ stiffens with increasing layer thickness\cite{LeeACS2010,LiAFM2012}, so that their frequency difference can be used to monitor the number of layers\cite{LeeACS2010}. The peak position difference$\sim$20cm$^{-1}$ is an indicator of 1L-MoS$_{2}$\cite{LeeACS2010}. The peak at$\sim417$cm$^{-1}$ (marked by asterisk in Fig.\ref{fig.3}a) corresponds to the A$_{1g}$ mode of sapphire\cite{PortJCP1967}.

The Raman spectrum measured at 514 nm of SLG on Cu is shown in Fig.\ref{fig.3}b (magenta curve). This is obtained after the removal of the non-flat background PL of Cu\cite{LagaAPL2013}. The two most intense features are the G and the 2D peak, with no significant D peak. The G peak corresponds to the E$_{2g}$ phonon at the Brillouin zone centre\cite{FerrNN2013}. The D peak is due to the breathing modes of sp$^2$ rings and requires a defect for its activation by double resonance\cite{FerrPRL2006,CancNL2011,FerrNN2013,FerrPRB2000}. The 2D peak is the second order of the D peak\cite{FerrNN2013}. This is always seen, even when no D peak is present, since no defects are required for the activation of two phonons with the same momentum, one backscattering from the other\cite{FerrNN2013}. In our sample, the 2D peak is a single sharp Lorentzian with FWHM(2D)$\sim$26cm$^{-1}$, a signature of SLG\cite{FerrPRL2006}. Different ($\sim$20) measurements show similar spectra, indicating uniform quality. The position of the G peak, Pos(G), is$\sim$1589cm$^{-1}$, with FWHM(G)$\sim$13cm$^{-1}$. The 2D peak position, Pos(2D) is$\sim$2698cm$^{-1}$, while the 2D to G peak intensity and area ratios, I(2D)/I(G) and A(2D)/A(G), are 2.6 and 5.8, respectively, indicating a p-doping$\sim$300meV\cite{DasNN2008, BrunACS2014, BaskPRB2009}, which corresponds to a carrier concentration$\sim$5$\cdot$10$^{12}$cm$^{-2}$.

\begin{figure}
\centerline{\includegraphics[width=90mm]{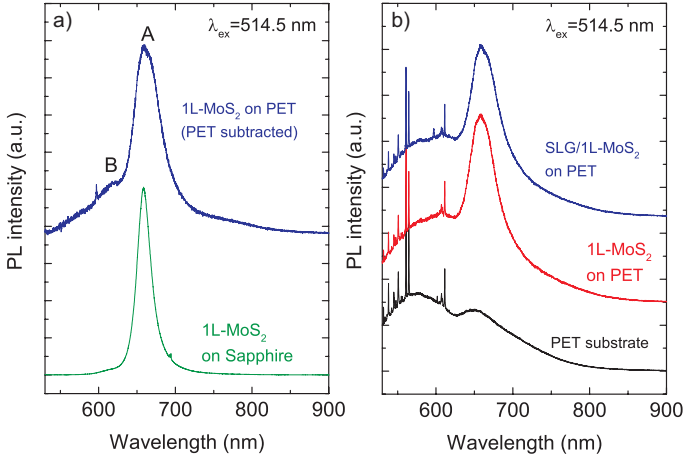}}
\caption{(a) PL spectrum at 514nm (2.41eV) of 1L-MoS$_{2}$ on sapphire, and SLG/1L-MoS$_2$ after transfer on PET. (b) PL spectra of PET substrate (black curve), 1L-MoS$_{2}$ on PET (red curve) and SLG/1L-MoS$_{2}$ on PET (blue curve).}
\label{fig.4}
\end{figure}

Another evidence for 1L-MoS$_{2}$ comes from the PL spectrum [Fig.\ref{fig.4}a (green curve)], showing a peak$\sim$658nm ($\sim$1.88eV), due to band-to-band radiative recombination of electron-hole pairs in 1L-MoS$_{2}$\cite{MakPRL2010}.

Then, the MoS$_2$ film is transferred onto a PET substrate from sapphire using a KOH-based approach\cite{DumcACS2015}. The samples are first spin coated with$\sim$100nm polymethyl methacrylate (PMMA). This is detached in a 30\% KOH solution, washed in DI water and transferred onto PET. The PMMA is then dissolved in acetone. Subsequently, SLG is transferred on the 1L-MoS$_2$ on PET. PMMA is spin coated on the SLG/Cu substrate, then placed in a solution of ammonium persulfate (APS) in DI water until Cu is etched\cite{BaeNN2010,BonaMT2012}. The PMMA membrane with attached SLG is then transferred to a beaker filled with DI water for cleaning APS residuals. The membrane is subsequently lifted with the target PET substrate having 1L-MoS$_2$ on top. After drying, PMMA is removed in acetone leaving SLG on 1L-MoS$_2$.

Raman and PL characterizations are performed at each step of the SLG/1L-MoS$_2$ assembly on PET, i.e on 1L-MoS$_2$ transferred on PET, and on SLG on 1L-MoS$_2$. This is to confirm no degradation during the fabrication process. For 1L-MoS$_2$ on PET, the Raman at 514nm is shown, with a close-up of the E$_{2g}^1$  and A$_{1g}$ regions, in Fig.\ref{fig.3}a (red curve). The frequency difference between E$_{2g}^1$ and A$_{1g}$ and the FWHMs are preserved on PET, suggesting no degradation. The PL spectrum of 1L-MoS$_{2}$ on PET is shown in Fig.\ref{fig.4}b (red curve). The signal from 1L-MoS$_{2}$ is convolved within the background due to the PET substrate [Fig.\ref{fig.4}b (black curve)]. In order to reveal the underlying PL signature of 1L-MoS$_{2}$, we use a point-to-point subtraction between the spectrum of 1L-MoS$_2$ on PET [Fig.\ref{fig.4}b (red curve)] and the reference PET spectrum [Fig.\ref{fig.4}b (black curve)]. Prior to subtraction, the spectra are normalized to the intensity of the Raman peak at$\sim$1615cm$^{-1}$ (corresponding to the peak at$\sim$560nm in Fig.\ref{fig.4}b), due to the stretching vibrations of benzene rings in PET\cite{BoerJPS1976}. As a result, the PL signal of 1L-MoS$_{2}$ can be seen in Fig.\ref{fig.4}a (blue curve) revealing no significant changes after transfer. The subsequent transfer of SLG on 1L-MoS$_{2}$ does not alter the 1L-MoS$_{2}$ PL position and lineshape [Fig.\ref{fig.4}b (blue curve)].

We then characterize the SLG transferred on 1L-MoS$_{2}$/PET. The intense Raman features of the underlying PET substrate\cite{BoerJPS1976} [Fig.\ref{fig.3}c (black curve)], mask the SLG peaks. In order to reveal the Raman signatures of SLG, we first measure the reference spectrum, shown in Fig.\ref{fig.3}c (black curve), of a PET substrate, using identical experimental conditions as those for SLG/1L-MoS$_{2}$/PET. We then implement a point-to-point subtraction, normalized to the intensity of the PET peak at $\sim$1615cm$^{-1}$, of the PET reference spectrum from the total spectrum Fig.\ref{fig.3}c (blue curve). The result is in Fig.\ref{fig.3}b (blue curve). The 2D peak retains its single-Lorentzian line-shape with FWHM(2D)$\sim$28cm$^{-1}$, validating the SLG transfer. The negligible D peak indicates that no significant defects are induced during transfer. Pos(G) is$\sim$1584cm$^{-1}$, FWHM(G)$\sim$15cm$^{-1}$, Pos(2D)$\sim$2685cm$^{-1}$, I(2D)/I(G)$\sim$2.9 and A(2D)/A(G)$\sim$5.9, indicating a p-doping$\sim$3$\cdot$10$^{12}$cm$^{-2}$ ($\sim$200meV)\cite{DasNN2008, BaskPRB2009}.
\begin{figure}
\centerline{\includegraphics[width=90mm]{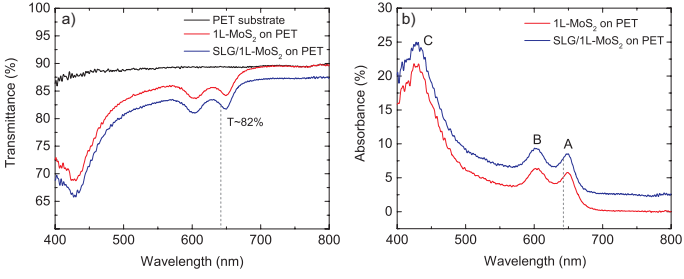}}
\caption{(a) Transmittance of PET (black curve), 1L-MoS$_2$ on PET (red curve) and SLG/1L-MoS$_2$ on PET (blue curve). (b) Absorbance of 1L-MoS$_2$ and SLG/1L-MoS$_2$ as derived from the transmittance measurements. Dashed lines indicate our test wavelength.}
\label{fig.5}
\end{figure}

We then measure the absorption and transmission of SLG/1L-MoS$_2$ using a broadband (400-1300nm) white light from a tungsten halogen lamp. The transmitted light is collected by a 10x objective lens (NA=0.25) with a Horiba Jobin Yvon HR800 spectrometer equipped with a 300 grooves/mm grating, charged coupled device (CCD) detector and a $50\mu$m pinhole. Fig.\ref{fig.5}a plots the optical transmittance of bare PET ($T_{PET}$, black line), 1L-MoS$_2$ on PET ($T_{MoS_2}$, red line) and the final SLG/1L-MoS$_2$ stack on PET ($T_{Hetero}$, blue line) measured in the 400-800nm wavelength range. Fig.\ref{fig.5}b plots the absorption of 1L-MoS$_2$ on PET ($Abs_{MoS_2}$, red line) and of SLG/1L-MoS$_2$ on PET ($Abs_{Hetero}$, blue line), calculated as $Abs_{MoS_2}$=($T_{PET}$-$T_{MoS_2}$)/$T_{PET}$ and $Abs_{Hetero}$=($T_{PET}$-$T_{Hetero}$)/$T_{PET}$. The three peaks in Fig.\ref{fig.5}b at $\sim$650nm (1.91eV), $\sim$603nm (2.06eV), and $\sim$428nm (2.90eV) correspond to the A, B, C excitons of 1L-MoS$_2$\cite{QiuPRL2013, MakPRL2010}. The positions of the A, B and C peaks remain unchanged after SLG transfer. The $Abs$ difference between the two curves (red and blue) is $\sim$2.6\%, consistent with the additional SLG absorption\cite{NairS2008}.

The PD area is shaped by etching, whereby SLG extending beyond the 1L-MoS$_2$ layer is removed in an oxygen plasma. The source-drain and gate electrodes are then defined by patterning the contacts area, followed by Cr/Au (6nm/60nm) evaporation and lift-off. PDs with different channels lengths (100$\mu$m-1mm), 2mm channel width and common side-gate electrodes (1cm x 0.5cm) are built, Fig.\ref{fig.1}b.

Refs.\citenum{RoyNN2013},\citenum{ZhanSR2014} showed that the responsivity of SLG/MoS$_2$ PDs can be enhanced by gating. This induces a stronger electric field at the SLG/MoS$_2$ interface and promotes charge transfer. Various gating techniques have been exploited for GRM-based devices, including conventional Si/SiO$_2$ back-gates\cite{NovoN2005}, high-k dielectrics (Al$_2$O$_3$, HfO$_2$)\cite{LemmEDL2007}, chemical dopants\cite{WehlACS2008}, ionic liquids\cite{YePNAS2011} and polymer electrolytes (PE)\cite{DasNN2008, BrunACS2014}. In order to gate our SLG/1L-MoS$_2$ on PET, we employ the latter due to its compatibility with flexible substrates\cite{SirrS2000} and the ability to substantially dope SLG ($\pm0.8eV$)\cite{DasNN2008, BrunACS2014} using small gate voltages (up to 4V), unlike other gating techniques, which would require considerably higher biases to reach the same doping\cite{NovoN2005, WehlACS2008}. We use a PE consisting of LiClO$_4$ and polyethylene oxide (POE)\cite{DasNN2008, BrunACS2014}. We place the PE over both the SLG channel and the side-gate electrode, and use $-1V<V_{GS}<1V$ in order to avoid electrochemical reactions, such as hydrolysis of residual water in the electrolyte\cite{AzaiJPS2007,EfetPRL2010}. These reactions may permanently modify the SLG electrode\cite{AzaiJPS2007,EfetPRL2010}, and compromise the stability and performance of the device.

We characterize the responsivity at 642nm ($\sim$1.93eV), slightly above the A exciton peak, where absorption of 1L-MoS$_2$ is maximized (Fig.\ref{fig.5}b). At 642nm the SLG/1L-MoS$_2$ heterostructure shows$\sim$8\% absorption (Fig.\ref{fig.5}b) and the device retains$\sim$82\% transparency (Fig.\ref{fig.5}a).

\begin{figure}
\centerline{\includegraphics[width=80mm]{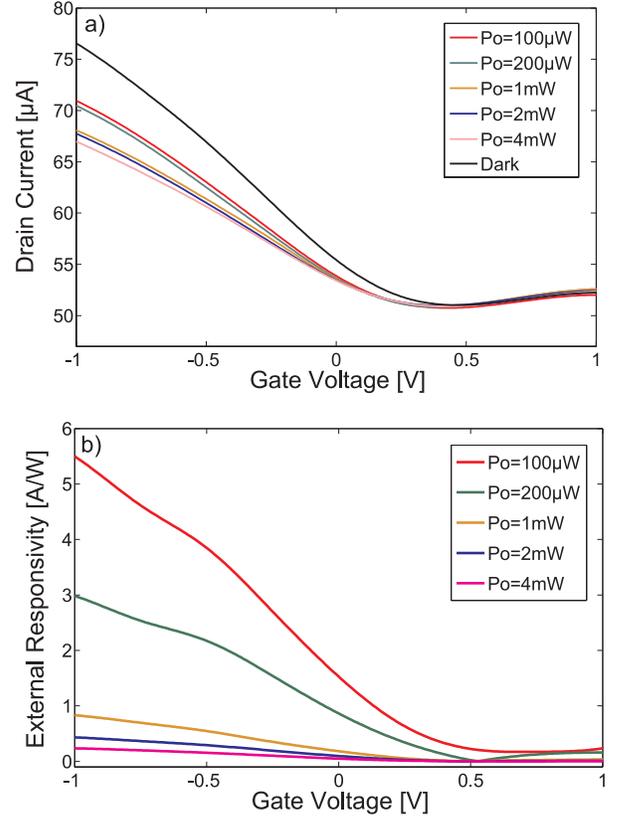}}
\caption{a) Transfer characteristics as a function of $P_o$. b) $R_{ext}$ as a function of $V_{GS}$ and $P_o$. Channel length and width are 100$\mu$m and 2mm respectively.}
\label{fig.6}
\end{figure}

The $I_{DS}-V_{GS}$ measurements in Fig.\ref{fig.6}a are done at room temperature using a probe station and a parameter analyzer (Keithley 4200). The PD is illuminated at normal incidence by a collimated laser with $P_o$ ranging from 100$\mu$W to 4mW. At these $P_o$ and with $V_{DS}=0.1V$ we measure a positive $V_{CNP}$ ranging from$\sim0.39V$ to $0.47V$, indicating an initial SLG p-doping$\sim$220meV, consistent with the Raman estimate.

Fig.\ref{fig.6}a shows that, for $-1V<V_{GS}<0.5V$ where SLG transport is hole dominated, the current decreases under illumination ($\sim10\mu A$ at $V_{GS}=-1V$), as anticipated from the band-diagram of Fig.\ref{fig.2}. For $V_{GS}>0.5V$, where SLG is electron-doped, the PD shows a small (up to$\sim0.2\mu A$) current increase under illumination. Fig.\ref{fig.6}b plots $R_{ext}$ as a function of $V_{GS}$, as derived from transconductance measurements using\cite{Sze2006}:
\begin{equation}
R_{ext}=\frac{I_{light}-I_{dark}} {P_o \cdot A_{PD}/A_o }
\label{eqn:1}
\end{equation}
where $I_{light}$ and $I_{dark}$ are the PD current under illumination and in dark, $I_{light}-I_{dark}=I_{ph}$, $A_o$ is the laser spot area, $A_{PD}$ is the PD area, and $A_{PD}/A_o$ is a scaling factor that takes into account the fact that only a fraction of optical power impinges the PD. As expected from the band-diagram in Fig.\ref{fig.2}, $R_{ext}$ tends to increase for more negative $V_{GS}$, up to$\sim5.5A/W$ at $V_{GS}=-1V$, $V_{DS}=0.1V$ for $P_o= 100\mu W$. By taking into account that only 8\% of light is absorbed ($P_{abs}=0.08\cdot P_o)$, we derive $R_{int}=R_{ext}/0.08=69A/W$. Fig.\ref{fig.6}b implies that the higher $P_o$, the lower $R_{ext}$. This can be explained considering that the more photo-generated electrons are injected into the p-doped channel, the lower the electric field at the SLG/1L-MoS$_2$ interface, therefore a reduced injection of electrons causes $R_{ext}$ to decrease.

\begin{figure}
\centerline{\includegraphics[width=80mm]{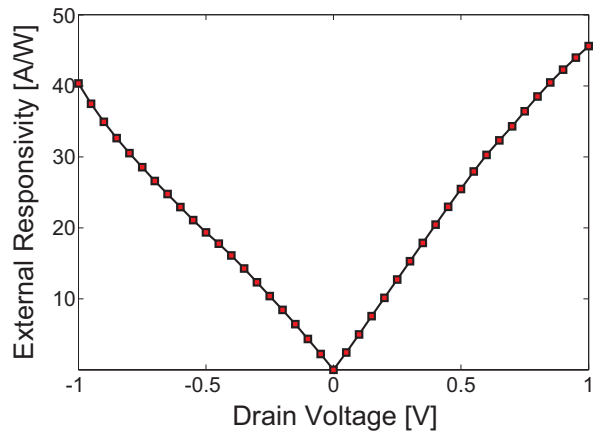}}
\caption{$R_{ext}$ as a function of $V_{DS}$ for $P_o=100\mu W$ at $V_{GS}=-1V$.}
\label{fig.7}
\end{figure}

Given that $R_{ext},R_{int}>1A/W$, we expect a photoconductive gain ($G_{PD}$)\cite{Sze2006, KonstNN2012}, whereby absorption of one photon results in multiple charge carriers contributing to I$_{ph}$. Our PDs act as optically-gated photoconductors, where the SLG conductance is modulated by optical absorption in the 1L-MoS$_2$. In this configuration, the presence of $G_{PD}$ implies that the injected electrons in SLG can recirculate multiple times between source and drain, before recombining with trapped holes in 1L-MoS$_2$. Consequently, $G_{PD}$ can be estimated as the ratio of electrons recombination ($\tau_{rec}$) and  transit ($t_{tr}$) times in the SLG channel: $G_{PD}=\tau_{rec}/t_{tr}$\cite{FerrN2015,KoppNN2014,Sze2006, KonstNN2012}. For higher $V_{DS}$, the free carriers drift velocity $\upsilon_{d}$ in the SLG channel increases linearly with bias (Ohmic region) until it saturates, because of carriers scattering with optical phonons\cite{MeriNN2008}. The linear increase in $\upsilon_{d}$ results in shorter $t_{tr}$, i.e $t_{tr}$ is defined as $L/\upsilon_{d}$, where $L$ is the channel length\cite{FerrN2015,KoppNN2014,Sze2006,KonstNN2012}. Therefore, $G_{PD}$ is also expected to grow linearly with $V_{DS}$, providing higher $R_{ext}$. To confirm the photoconductive nature of $G_{PD}$ in our device and test the dependence of $R_{ext}$ on $V_{DS}$, we measure $I_{DS}-V_{DS}$ under illumination at $P_o=100 \mu W$ for $V_{GS}=-1V$ and calculate $R_{ext}$ using Eq.\ref{eqn:1}. We use $V_{DS}<1V$ to keep the device operation in the linear (Ohmic) regime and minimize effect of non-linear dependence of $\upsilon_{d}$ on $V_{DS}$ (such as velocity saturation) that might appear for $V_{DS}>1V$\cite{MeriNN2008}. As shown in Fig.\ref{fig.7}, $R_{ext}$ scales with $V_{DS}$ and reaches $\sim45.5A/W$ ($R_{int}\sim 570A/W$) at $V_{DS}=1V$. This is almost one order of magnitude higher than at $V_{DS}=0.1V$, consistent with the similar increase in $V_{DS}$. These results are at least two orders of magnitude higher than semiconductor flexible membranes\cite{YangAPL2010,YuanAPL2009} and five orders of magnitude larger than other flexible PDs based on GRMs\cite{WithNL2014,FinnJMCC2014}.

We define $G_{PD}$ as the ratio between electrons recirculating in the SLG channel, thus sustaining $I_{ph}$, and the initial electron concentration injected into SLG from 1L-MoS$_2$\cite{ZhanSR2014}:
\begin{equation}
G_{PD}=\frac{I_{light}-I_{dark}} {q\cdot A_{PD}\cdot \Delta n_{ch} }
\label{eqn:2}
\end{equation}
where $q$ is the electron charge and $\Delta n_{ch}$ is the concentration per unit area of the injected electrons. $\Delta n_{ch}$ is equal to the trapped-hole concentration in 1L-MoS$_2$, which is related to charge neutrality point shift $\Delta V_{GS}=\Delta V_{CNP}$ in the transfer characteristics. To calculate $\Delta n_{ch}$, we first write the potential balance in the metal-dielectric-SLG structure. When $V_G$ is applied, it creates a gate-to-channel potential drop ($V_{diel}$), and it induces a local electrostatic potential in graphene channel ($V_{ch}=E_{F}/q$)\cite{Sze2006, DasNN2008}:
\begin{equation}
V_G=V_{diel}+V_{ch}=\frac {Q_G}{C_G}+V_{ch}
\label{eqn:3}
\end{equation}
where $Q_G$ and $C_G$ are the charge concentration and the geometrical capacitance per unit area associated with the gate electrode respectively. $|Q_G|=|q\cdot n_{ch}|$, reflecting the charge neutrality of the gate capacitor, with $n_{ch}$ the charge carrier concentration per unit area in the channel. Any variations $\Delta n_{ch}$ change $\Delta V_G$. As a result:
\begin{equation}
\frac {dV_G}{dQ_G}=\frac {1}{C_G}+\frac {dV_{ch}}{dQ_G}
\label{eqn:4}
\end{equation}
which leads to:
\begin{equation}
\Delta Q_G=(1/C_G+1/C_Q)^{-1}\cdot \Delta V_G
\label{eqn:5}
\end{equation}
where $C_Q=dQ_{G}/dV_{ch}$ is the SLG quantum capacitance\cite{FangAPL2007,XiaNN2009-2} that characterizes the changes of the channel potential $\Delta V_{ch}$ as a result of additional gating $\Delta Q_{G}$, and $(1/C_G+1/C_Q)^{-1}$ is the total capacitance $C_{tot}$.

\begin{figure}
\centerline{\includegraphics[width=60mm]{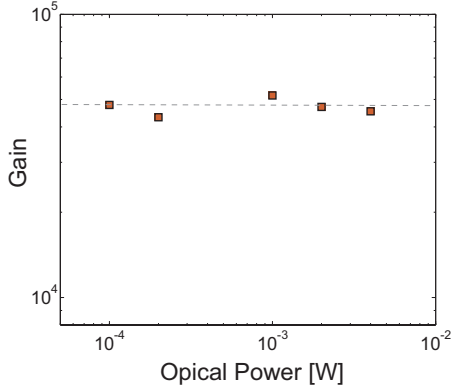}}
\caption{$G_{PD}$ as a function of $P_o$ at $V_{GS}=-1V$ and $V_{DS}=0.1V$.}
\label{fig.8}
\end{figure}

To calculate $Q_G$ we first need to find $C_G$ and $C_Q$. In PE gating, $C_G$ is associated with the EDL at the SLG/electrolyte interface\cite{DasNN2008,LuNL2004,OzelNL2005, XiaNN2009-2}. The EDL acts like a parallel-plate capacitor with an dielectric layer thickness of the order of $\lambda_D$, so that $C_G=C_{EDL}=\epsilon \epsilon_0/ \lambda_D$, where $\epsilon$ is the PE dielectric constant, and $\epsilon_0$ is the vacuum permittivity. In principle, for a monovalent electrolyte, $\lambda_D$ can be explicitly calculated\cite{Russ1989} if the electrolyte concentration is known. However, in the presence of a polymer matrix, the electrolyte ions can form complexes with polymer chains\cite{Salomon1994}, therefore the precise ion concentration is difficult to measure. For PE gating, different EDL thicknesses in the range$\sim1-5nm$ have been reported\cite{DasPRB2009,DasNN2008, LuNL2004, OzelNL2005}. To estimate $C_{EDL}$ in our devices we take $\lambda_D\sim2nm$\cite{DasNN2008} and the dielectric constant of the PEO matrix to be $\epsilon\sim5$\cite{Boyd1983}, as done in Ref.\citenum{DasNN2008}. As a result, we obtain $C_{EDL}=2.2\times10^{-6}F/cm^{2}$. This is the same order of magnitude as the SLG $C_Q$\cite{XiaNN2009-2}. Therefore the latter cannot be neglected in Eq.\ref{eqn:5}. $C_Q$ is given by\cite{XiaNN2009-2}:
\begin{equation}
C_Q\approx\frac{2q^2}{\hbar v_F \sqrt{\pi}}\cdot \sqrt{n_{ch}+n_i}
\label{eqn:7}
\end{equation}
where $\hbar$ is the reduced Planck constant, $v_F=1.1\cdot10^6m/s$ is the Fermi velocity of charge carriers in graphene\cite{NovoN2005, ZhanN2005} and $n_i$ is the intrinsic carrier concentration in SLG near the Dirac point induced by charge impurities, defects and local potential fluctuations in the SLG channel\cite{ChenNP2008, AdamPNAS2007, VictorPRB2007, XiaNN2009-2}. Using Raman and transconductance we estimate $n_i\sim3\cdot10^{12}cm^{-2}$. From Eq.\ref{eqn:7} we then get $C_Q=4\cdot 10^{-6}F/cm^{2}$ at $V_{CNP}$. From Fig.\ref{fig.6}a, and extracting $\Delta V_{CNP}$ between the dark current and the transfer curves measured under illumination, and with Eq.\ref{eqn:5}, we get $\Delta n_{ch}$ ranging from $4-8\cdot10^{11}cm^{-2}$ for $P_o$ going from $100\mu W$ to $4mW$. As a result, we obtain $G_{PD}\sim5\times 10^4$ at $V_{DS}=0.1V$ for different $P_o$ as shown in Fig.\ref{fig.8}. As discussed previously, $G_{PD}$ becomes larger for higher $V_{DS}$. Thus, we measure an increase of almost order of magnitude ($G_{PD}\sim4\cdot10^5$ at $P_o=100\mu W$) for $V_{DS}$ going from $0.1V$ to $1V$.
\begin{figure}
\centerline{\includegraphics[width=90mm]{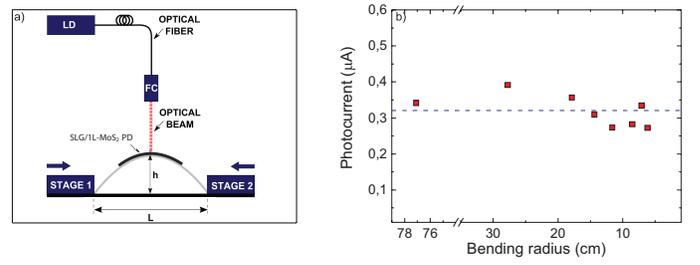}}
\caption{a) Schematic two-points bending setup. LD = laser diode;  FC= fiber collimator; b) I$_{ph}$ as a function of r$_b$. The dashed line shows the mean value.}
\label{fig.9}
\end{figure}

Finally, we test I$_{ph}$ as a function of bending. $r_b$ is estimated as $(h^2+(l/2)^2)/2h$, where $l$ is the chord of circumference connecting the two ends of the arc, and $h$ is the height at the chord midpoint. Fig.\ref{fig.9}b plots $I_{ph}$ for different $r_b$, showing a maximal deviation of $15\%$ for $r_b$ down to $6cm$. This value is comparable to r$_b$ reported for semiconductor membrane PDs\cite{YangAPL2010,YuanAPL2009}, yet the latter show two orders of magnitude lower ($<0.3A/W$) responsivities\cite{YangAPL2010,YuanAPL2009}. Although our r$_b$ is one order of magnitude larger than reported by flexible single NW devices\cite{LeeMEMS2012, LiuOE2013, ChenJMCC2014, YuJMCC2014}, the device area of our PDs ($>0.2mm^2$) is at least four orders of magnitude larger compared to these devices ($<5\mu m^2$).
\subsection{Conclusions}
We reported flexible PDs for visible wavelengths with external responsivity up to$\sim$45.5A/W and photoconductive gain of $4\times10^5$. This is at least two orders of magnitude higher than semiconductor flexible membranes and other GRM-based flexible PDs reported thus far. Our PDs show stable performance upon bending for radii of curvature larger than $\sim$6cm. The responsivity, flexibility, transparency and low operation voltage (below $1V$) of our PDs makes them attractive candidates for wearable, biomedical and low-power opto-electronic applications.

We acknowledge funding from EU Graphene Flagship (no. 604391), ERC Grant Hetero2D, EPSRC Grants EP/K01711X/1, EP/K017144/1, EU grant GENIUS, a Royal Society Wolfson Research Merit Award


\begin{thebibliography}{100}
\bibitem{AkinNC2014} Akinwande, D., Petrone, N., Hone, J. \emph{Nat. Commun.} \textbf {2014}, \emph{5}.
\bibitem{Ryha2010} Ryhaenen, T. T., \textbf{2010}, Cambridge University Press.
\bibitem{LiuOE2013} Liu, Z., Chen, G., Liang, B., Yu, G., Huang, H., Chen, D., Shen, G. \emph{Opt. Express} \textbf {2013}, \emph{21}, 7799-7810.
\bibitem{YuanAPL2009} Yuan, H.-C., Shin, J., Qin, G., Sun, L., Bhattacharya, P., Lagally, M. G., Celler, G. K., Ma, Z. \emph{Appl. Phys. Lett.} \textbf {2009}, \emph{94}, 13102.
\bibitem{ParkS2009} Park, S.-I., Xiong, Y., Kim, R.-H., Elvikis, P., Meitl, M., Kim, D.-H., Wu, J., Yoon, J., Yu, C.-J., Liu, Z., et al. \emph{Science} \textbf {2009}, \emph{325}, 977-981.
\bibitem{QianAPL2008} Qiang, Z., Yang, H., Chen, L., Pang, H., Ma, Z., Zhou, W. \emph{App. Phys. Lett.} \textbf {2008}, \emph{93}, 061106.
\bibitem{BosmPTL2010} Bosman, E., Van Steenberge, G., Van Hoe, B., Missinne, J., Vanfleteren, J., Van Daele, P. \emph{Photon. Technol. Lett., IEEE} \textbf {2010}, \emph{22}, 287-289.
\bibitem{ChenNM2011} Chen, Z., Ren, W., Gao, L., Liu, B., Pei, S., Cheng, H.-M. \emph{Nat. Mater.} \textbf {2011}, \emph{10}, 424-428.
\bibitem{ShahNP2010} Shahi, S. \emph{Nature Photon.} \textbf {2010}, \emph{4}, 506-506.
\bibitem{YoonNC2011} Yoon, J., Li, L., Semichaevsky, A. V., Ryu, J. H., Johnson, H. T., Nuzzo, R. G., Rogers, J. A. \emph{Nat. Commun.} \textbf {2011}, \emph{2}, 343.
\bibitem{KimS2011} Kim, D.-H., Lu, N., Ma, R., Kim, Y.-S., Kim, R.-H., Wang, S., Wu, J., Won, S. M., Tao, H., Islam, A., et al. \emph{Science} \textbf {2011}, \emph{333}, 838-843.
\bibitem{KoN2008} Ko, H. C., Stoykovich, M. P., Song, J., Malyarchuk, V., Choi, W. M., Yu, C.-J., Geddes Iii, J. B., Xiao, J., Wang, S., Huang, Y., et al. \emph{Nature} \textbf {2008}, \emph{454}, 748-753.
\bibitem{BlakJAP1982} Blakemore, J. S. \emph{J. Appl. Phys.} \textbf {1982}, \emph{53}, 123-181.
\bibitem{MacmJMS1972} MacMillan, N. H. \emph{J Mater. Sci.} \textbf {1972}, \emph{7}, 239-254.
\bibitem{YangAPL2010} Yang, W., Yang, H., Qin, G., Ma, Z., Berggren, J., Hammar, M., Soref, R., Zhou, W. \emph{Appl. Phys. Lett.} \textbf {2010}, \emph{96}, 121107.
\bibitem{ChenJMCC2014} Chen, G., Liang, B., Liu, Z., Yu, G., Xie, X., Luo, T., Xie, Z., Chen, D., Zhu, M.-Q., Shen, G. \emph{J. Mater. Chem. C} \textbf {2014}, \emph{2}, 1270-1277.
\bibitem{LeeMEMS2012} Lee, S., Jung, S. W., Park, S., Ahn, J., Hong, S. J., Yoo, H. J., Lee, M. H., Cho, D. I. \emph{Proc. IEEE Micr. Elect.}, \textbf {2012}.
\bibitem{YuJMCC2014} Yu, G., Liu, Z., Xie, X., Ouyang, X., Shen, G. \emph{J. Mater. Chem. C} \textbf {2014}, \emph{2}, 6104-6110.
\bibitem{Sze2006} Sze, S. M., Kwok, K. N. \textbf {2006}, Wiley, New York.
\bibitem{BonaNP2010} Bonaccorso, F., Sun, Z., Hasan, T., Ferrari, A. C. \emph{Nature Photon.} \textbf {2010}, \emph{4}, 611-622.
\bibitem{FerrN2015} Ferrari, A. C., Bonaccorso, F., Fal'ko, V., Novoselov, K. S., Roche, S., Boggild, P., Borini, S., Koppens, F. H. L., Palermo, V., Pugno, N., et al. \emph{Nanoscale} \textbf {2015}, \emph{7}, 4598-4810.
\bibitem{KoppNN2014} Koppens, F. H. L., Mueller, T., Avouris, P., Ferrari, A. C., Vitiello, M. S., Polini, M. \emph{Nat. Nanotechnol.} \textbf {2014}, \emph{9}, 780-793.
\bibitem{SunACS2010} Sun, Z. P., Hasan, T., Torrisi, F., Popa, D., Privitera, G., Wang, F. Q., Bonaccorso, F., Basko, D. M., Ferrari, A. C. \emph{Acs Nano} \textbf {2010}, \emph{4}, 803-810.
\bibitem{KimN2009} Kim, K. S., Zhao, Y., Jang, H., Lee, S. Y., Kim, J. M., Kim, K. S., Ahn, J.-H., Kim, P., Choi, J.-Y., Hong, B. H. \emph{Nature} \textbf {2009}, \emph{457}, 706-710.
\bibitem{BaugNN2014} Baugher, B. W. H., Churchill, H. O. H., Yang, Y., Jarillo-Herrero, P. \emph{Nat. Nanotechnol.} \textbf {2014}, \emph{9}, 262-267.
\bibitem{PospNN2014} Pospischil, A., Furchi, M. M., Mueller, T. \emph{Nat. Nanotechnol.} \textbf {2014}, \emph{9}, 257-261.
\bibitem{KisNN2013} Lopez-Sanchez, O.; Lembke, D.; Kayci, M.; Radenovic, A.; Kis, A. \emph{Nat. Nanotechnol.} \textbf {2013}, \emph{8},497–501.
\bibitem{XiaNN2009} Xia, F.; Mueller, T.; Lin, Y. M.; Valdes-Garcia, A.; Avouris, P. \emph{Nat. Nanotechnol.} \textbf {2009}, \emph{4}, 839-843.
\bibitem{LiuN2011} Liu, M., Yin, X. B., Ulin-Avila, E., Geng, B. S., Zentgraf, T., Ju, L., Wang, F., Zhang, X. \emph{Nature} \textbf {2011}, \emph{474}, 64-67.
\bibitem{ChenN2012} Chen, J., Badioli, M., Alonso-Gonzalez, P., Thongrattanasiri, S., Huth, F., Osmond, J., Spasenovic, M., Centeno, A., Pesquera, A., Godignon, P., et al. \emph{Nature} \textbf {2012}, \emph{487}, 77-81.
\bibitem{EchtNC2011} Echtermeyer, T. J., Britnell, L., Jasnos, P. K., Lombardo, A., Gorbachev, R. V., Grigorenko, A. N., Geim, A. K., Ferrari, A. C., Novoselov, K. S. \emph{Nat. Commun.} \textbf {2011}, \emph{2}, 458.
\bibitem{FeiN2012} Fei, Z., Rodin, A. S., Andreev, G. O., Bao, W., McLeod, A. S., Wagner, M., Zhang, L. M., Zhao, Z., Thiemens, M., Dominguez, G., et al. \emph{Nature} \textbf {2012}, \emph{487}, 82-85.
\bibitem{JuNN2011} Ju, L., Geng, B., Horng, J., Girit, C., Martin, M., Hao, Z., Bechtel, H. A., Liang, X., Zettl, A., Shen, Y. R., et al. \emph{Nat. Nanotechnol.} \textbf {2011}, \emph{6}, 630-634.
\bibitem{YanNN2012} Yan, H., Li, X., Chandra, B., Tulevski, G., Wu, Y., Freitag, M., Zhu, W., Avouris, P., Xia, F. \emph{Nat. Nanotechnol.} \textbf {2012}, \emph{7}, 330-334.
\bibitem{FurcNL2014} Furchi, M. M., Pospischil, A., Libisch, F., Burgdörfer, J., Mueller, T. \emph{Nano Lett.} \textbf {2014}, \emph{14}, 4785-4791.
\bibitem{WithNM2015} Withers, F., Del Pozo-Zamudio, O., Mishchenko, A., Rooney, A. P., Gholinia, A., Watanabe, K., Taniguchi, T., Haigh, S. J., Geim, A. K., Tartakovskii, A. I., et al. \emph{Nat. Mater.} \textbf {2015}, \emph{14}, 301-306.
\bibitem{RoyNN2013} Roy, K., Padmanabhan, M., Goswami, S., Sai, T. P., Ramalingam, G., Raghavan, S., Ghosh, A. \emph{Nat. Nanotechnol.} \textbf {2013}, \emph{8}, 826-830.
\bibitem{ZhanSR2014} Zhang, W. J., Chuu, C. P., Huang, J. K., Chen, C. H., Tsai, M. L., Chang, Y. H., Liang, C. T., Chen, Y. Z., Chueh, Y. L., He, J. H., et al. \emph{Sci. Rep.} \textbf {2014}, \emph{4}.
\bibitem{FinnJMCC2014} Finn, D. J., Lotya, M., Cunningham, G., Smith, R. J., McCloskey, D., Donegan, J. F., Coleman, J. N. \emph{J. Mater. Chem. C} \textbf {2014}, \emph{2}, 925-932.
\bibitem{WithNL2014} Withers, F., Yang, H., Britnell, L., Rooney, A. P., Lewis, E., Felten, A., Woods, C. R., Romaguera, V. S., Georgiou, T., Eckmann, A., et al. \emph{Nano Lett.} \textbf {2014}, \emph{14}, 3987-3992.
\bibitem{BonaMT2012} Bonaccorso, F., Lombardo, A., Hasan, T., Sun, Z. P., Colombo, L., Ferrari, A. C. \emph{Mater. Today} \textbf {2012}, \emph{15}, 564-589.
\bibitem{HernNN2008} Hernandez, Y., Nicolosi, V., Lotya, M., Blighe, F. M., Sun, Z. Y., De, S., McGovern, I. T., Holland, B., Byrne, M., Gun'ko, Y. K. et al. \emph{Nat. Nanotechnol.} \textbf {2008}, \emph{3}, 563-568.
\bibitem{KingACS2010} King, P. J., Khan, U., Lotya, M., De, S., Coleman, J. N. \emph{Acs Nano} \textbf {2010}, \emph{4}, 4238-4246.
\bibitem{TorrNN2014} Torrisi, F., Coleman, J. N. \emph{Nat. Nanotechnol.} \textbf {2014}, \emph{9}, 738-739.
\bibitem{FaraOAM2011} Faraj, M. G., Ibrahim, K., Ali, M. K. M. \emph{Optoelectron Adv. Mat.} \textbf {2011}, \emph{5}, 879-882.
\bibitem{MartPNAS2013} Martins, L. G. P., Song, Y., Zeng, T. Y., Dresselhaus, M. S., Kong, J., Araujo, P. T. \emph{Proc. Natl Acad. Sci. USA} \textbf {2013}, \emph{110}, 17762-17767.
\bibitem{BaeNN2010} Bae, S., Kim, H., Lee, Y., Xu, X., Park, J. S., Zheng, Y., Balakrishnan, J., Lei, T., Kim, H. R., Song, Y. I., et al. \emph{Nat. Nanotechnol.} \textbf {2010}, \emph{5}, 574-8.
\bibitem{DasNN2008} Das, A., Pisana, S., Chakraborty, B., Piscanec, S., Saha, S. K., Waghmare, U. V., Novoselov, K. S., Krishnamurthy, H. R., Geim, A. K., Ferrari, A. C., Sood, A. K. \emph{Nat. Nanotechnol.} \textbf {2008}, \emph{3}, 210-5.
\bibitem{DasPRB2009} Das, A., Chakraborty, B., Piscanec, S., Pisana, S., Sood, A. K., Ferrari, A. C. \emph{Phys. Rev. B} \textbf {2009}, \emph{79}, 155417.
\bibitem{ChoiNC2013} Choi, M. S., Lee, G. H., Yu, Y. J., Lee, D. Y., Lee, S. H., Kim, P., Hone, J., Yoo, W. J., \emph{Nat. Commun.} \textbf {2013}, \emph{4}, 1624.
\bibitem{DasNL2013} Das, S., Chen, H. Y., Penumatcha, A. V., Appenzeller, J., \emph{Nano Lett.} \textbf {2013}, \emph{13}, 100-105.
\bibitem{ShanPRL2005} Shan, B., Cho, K. \emph{Phys. Rev. Lett.} \textbf {2005}, \emph{94}, 236602.
\bibitem{YuNL2009}  Yu, Y.-J., Zhao, Y., Ryu, S., Brus, L. E., Kim, K. S., Kim, P. \emph{Nano Lett.} \textbf {2009}, \emph{9}, 3430-3434.
\bibitem{KimNL2010} Kim, B. J., Jang, H., Lee, S.-K., Hong, B. H., Ahn, J.-H., Cho, J. H. \emph{Nano Lett.} \textbf {2010}, \emph{10}, 3464-3466.
\bibitem{LeeNL2012} Lee, S. K., Jang, H. Y., Jang, S., Choi, E., Hong, B. H., Lee, J., Park, S., Ahn, J. H. \emph{Nano Lett.} \textbf {2012}, \emph{12}, 3472-3476.
    \bibitem{DumcACS2015} Dumcenco, D., Ovchinnikov, D., Marinov, K., Lazic, P., Gibertini, M., Marzari, N., Sanchez, O. L., Kung, Y.-C., Krasnozhon, D., Chen, M.-W. et al. \emph{Acs Nano} \textbf {2015}, \emph{9}, 4611-4620.
\bibitem{LiS2009} Li, X. S., Cai, W. W., An, J. H., Kim, S., Nah, J., Yang, D. X., Piner, R., Velamakanni, A., Jung, I., Tutuc, E., et al. \emph{Science} \textbf {2009}, \emph{324}, 1312-1314.
\bibitem{VerbPRL1970} Verble, J. L., Wieting, T. J. \emph{Phys. Rev. Lett.} \textbf {1970}, \emph{25}, 362-365.
\bibitem{WietPRB1971} Wieting, T. J., Verble, J. L. \emph{Phys. Rev. B} \textbf {1971}, \emph{3}, 4286-4292.
\bibitem{LeeACS2010} Lee, C., Yan, H. G., Brus, L. E., Heinz, T. F., Hone, J., Ryu, S. \emph{Acs Nano} \textbf {2010}, \emph{4}, 2695-2700.
\bibitem{LiAFM2012} Li, H., Zhang, Q., Yap, C. C. R., Tay, B. K., Edwin, T. H. T., Olivier, A., Baillargeat, D. \emph{Adv. Funct. Mater.} \textbf {2012}, \emph{22}, 1385-1390.
\bibitem{PortJCP1967} Porto, S. P. S., Krishnan, R. S. \emph{J. Chem. Phys.} \textbf {1967}, \emph{47}, 1009-1012.
\bibitem{LagaAPL2013} Lagatsky, A. A., Sun, Z., Kulmala, T. S., Sundaram, R. S., Milana, S., Torrisi, F., Antipov, O. L., Lee, Y., Ahn, J. H., Brown, C. T. A., et al. \emph{Appl. Phys. Lett.} \textbf {2013}, \emph{102}, 013113.
\bibitem{FerrPRL2006} Ferrari, A. C., Meyer, J. C., Scardaci, V., Casiraghi, C., Lazzeri, M., Mauri, F., Piscanec, S., Jiang, D., Novoselov, K. S., Roth, S., et al. \emph{Phys. Rev. Lett.} \textbf {2006}, \emph{97}, 187401.
\bibitem{CancNL2011} Cancado, L. G., Jorio, A., Ferreira, E. H., Stavale, F., Achete, C. A., Capaz, R. B., Moutinho, M. V., Lombardo, A., Kulmala, T. S., Ferrari, A. C. \emph{Nano Lett.} \textbf {2011}, \emph{11}, 3190-6.
\bibitem{FerrNN2013} Ferrari, A. C., Basko, D. M. \emph{Nat. Nanotechnol.} \textbf {2013}, \emph{8}, 235-46.
\bibitem{FerrPRB2000} Ferrari, A. C., Robertson, J. \emph{Phys. Rev. B} \textbf {2000}, \emph{61}, 14095-14107.
\bibitem{BaskPRB2009} Basko, D. M., Piscanec, S., Ferrari, A. C. \emph{Phys. Rev. B} \textbf {2009}, \emph{80}, 165413.
\bibitem{BrunACS2014} Bruna, M., Ott, A. K., Ijas, M., Yoon, D., Sassi, U., Ferrari, A. C. \emph{Acs Nano} \textbf {2014}, \emph{8}, 7432-7441.
\bibitem{MakPRL2010} Mak, K. F., Lee, C., Hone, J., Shan, J., Heinz, T. F. \emph{Phys. Rev. Lett.} \textbf {2010}, \emph{105}, 136805.
\bibitem{BoerJPS1976} Boerio, F. J., Bahl, S. K., McGraw, G. E. \emph{J. Poly. Sci.} \textbf {1976}, \emph{14}, 1029-1046.
\bibitem{QiuPRL2013} Qiu, D. Y., da Jornada, F. H., Louie, S. G. \emph{Phys. Rev. Lett.} \textbf {2013}, \emph{111}, 216805.
\bibitem{NairS2008} Nair, R. R., Blake, P., Grigorenko, A. N., Novoselov, K. S., Booth, T. J., Stauber, T., Peres, N. M. R., Geim, A. K. \emph{Science} \textbf {2008}, \emph{320}, 1308-1308.
\bibitem{NovoN2005} Novoselov, K. S., Geim, A. K., Morozov, S. V., Jiang, D., Katsnelson, M. I., Grigorieva, I. V., Dubonos, S. V., Firsov, A. A. \emph{Nature} \textbf {2005}, \emph{438}, 197-200.
\bibitem{LemmEDL2007} Lemme, M. C., Echtermeyer, T. J., Baus, M., Kurz, H. \emph{IEEE Electr. Device L.} \textbf {2007}, \emph{28}, 282-284.
\bibitem{WehlACS2008} Wehling, T. O., Novoselov, K. S., Morozov, S. V., Vdovin, E. E., Katsnelson, M. I., Geim, A. K., Lichtenstein, A. I. \emph{Nano Lett.} \textbf {2008}, \emph{8}, 173-177.
\bibitem{YePNAS2011} Ye, J. T., Craciun, M. F., Koshino, M., Russo, S., Inoue, S., Yuan, H. T., Shimotani, H., Morpurgo, A. F., Iwasa, Y. \emph{Proc. Natl Acad. Sci. USA} \textbf {2011}, \emph{108}, 13002-13006.
\bibitem{SirrS2000} Sirringhaus, H., Kawase, T., Friend, R. H., Shimoda, T., Inbasekaran, M., Wu, W., Woo, E. P. \emph{Science} \textbf {2000}, \emph{290}, 2123-2126.
\bibitem{AzaiJPS2007} Azais, P., Duclaux, L., Florian, P., Massiot, D., Lillo-Rodenas, M.-A., Linares-Solano, A., Peres, J.-P., Jehoulet, C., Béguin, F. \emph{J. Power Sources} \textbf {2007}, \emph{171}, 1046-1053.
\bibitem{EfetPRL2010} Efetov, D. K., Kim, P. \emph{Phys. Rev. Lett.} \textbf {2010}, \emph{105}, 256805.
\bibitem{KonstNN2012} Konstantatos, G., Badioli, M., Gaudreau, L., Osmond, J., Bernechea, M., de Arquer, G., Gatti, F., Koppens, F. H. L. \emph{Nat. Nanotechnol.} \textbf {2012}, \emph{7}, 363-368.
\bibitem{MeriNN2008} Meric, I., Han, M. Y., Young, A. F., Ozyilmaz, B., Kim, P., Shepard, K. L. \emph{Nat. Nanotechnol.} \textbf {2008}, \emph{3}, 654-659.
\bibitem{XiaNN2009-2} Xia, J., Chen, F., Li, J., Tao, N. \emph{Nat. Nanotechnol.} \textbf {2009}, \emph{4}, 505-509.
\bibitem{FangAPL2007} Fang, T., Konar, A., Xing, H., Jena, D. \emph{Appl. Phys. Lett.} \textbf {2007}, \emph{9}, 092109.
\bibitem{OzelNL2005} Ozel, T., Gaur, A., Rogers, J. A., Shim, M. \emph{Nano Lett.} \textbf {2005}, \emph{5}, 905-911.
\bibitem{LuNL2004} Lu, C., Fu, Q., Huang, S., Liu, J. \emph{Nano Lett.} \textbf {2004}, \emph{4}, 623-627.
\bibitem{Russ1989} Russel, W.B., Saville, D.A. and Schowalter, W. R. \textbf {1989}, Cambridge University Press, UK.
\bibitem{Salomon1994} Salomon, M., Xu, M., Eyring, E. M., Petrucci, S. \emph{J. Phys. Chem.} \textbf {1994}, \emph{98}, 8234–8244.
\bibitem{Boyd1983} Boyd, R. H. \emph{J. Polym. Sci. Polym. Phys.} \textbf {1983}, \emph{21}, 505–514.
\bibitem{ZhanN2005} Zhang, Y. B., Tan, Y. W., Stormer, H. L., Kim, P. \emph{Nature} \textbf {2005}, \emph{438}, 201-204.
\bibitem{AdamPNAS2007} Adam, S., Hwang, E. H., Galitski, V. M., Sarma, S. D. \emph{Proc. Natl. Acad. Sci. USA} \textbf {2007}, \emph{104}, 18392-18397.
\bibitem{ChenNP2008} Chen, J. H.,  Jang, C.,  Adam, S.,  Fuhrer, M. S.,  Williams, E. D., Ishigami, M. \emph{Nature Phys.} \textbf {2008}, \emph{4}, 377-381.
\bibitem{VictorPRB2007} Victor, M. G., Shaffique, A., Sarma, S. D. \emph{Phys. Rev. B} \textbf {2007}, \emph{76}, 245405.
\end{thebibliography}
\end{document}